 \documentclass[manuscript]{aastex63}

\usepackage{amssymb}	
\usepackage{amsmath}	
\usepackage{bm}		
\received{June 13, 2020}
\revised{??, 2020}
\accepted{\today}
\submitjournal{ApJ}

\shorttitle{Peak energy of prompt $\gamma$-rays in GRBs}
\shortauthors{Zhang et al.}

\begin{document}

\title{On Spectral Peak Energy of \textit{Swift} Gamma-Ray Bursts}

\correspondingauthor{Z. B. Zhang}
\email{z\_b\_zhang@sina.com}

\author{Z. B. Zhang }
\affiliation{Department of Physics, College of Physics, Guizhou University, Guiyang 550025, P. R. China}
\affiliation{College of Physics and Engineering, Qufu Normal University, Qufu 273165, P. R. China}
\author{M. Jiang}
\affiliation{Department of Physics, College of Physics, Guizhou University, Guiyang 550025, P. R. China}
\author{Y. Zhang}
\affiliation{College of Physics and Engineering, Qufu Normal University, Qufu 273165, P. R. China}
\author{K. Zhang}
\affiliation{College of Physics and Engineering, Qufu Normal University, Qufu 273165, P. R. China}
\author{X. J. Li}
\affiliation{College of Physics and Engineering, Qufu Normal University, Qufu 273165, P. R. China}
\author{Q. Zhang}
\affiliation{Department of Physics, College of Physics, Guizhou University, Guiyang 550025, P. R. China}
%



\begin{abstract}

Owing to narrow energy band of \textit{Swift}/BAT, several urgent issues are required to pay more attentions but unsolved so far. We systematically study the properties of a refined sample of 283 \textit{Swift}/BAT gamma-ray bursts with well-measured spectral peak energy ($E_{\text p}$) at a high confidence level larger than 3$\sigma$. It is interestingly found that duration ($T_{90}$) distribution of \textit{Swift} bursts still exhibits an evident bimodality with a more reliable boundary of $T_{90}\simeq$1.06 s instead of 2 s for previously contaminated samples including bursts without well-peaked spectra, which is very close to $\sim$1.27 s and $\sim$0.8 s suggested by some authors for Fermi/GBM and \textit{Swift}/BAT catalogs, respectively. The \textit{Swift}/BAT short and long bursts have comparable mean $E_{\text p}$ values of $87^{+112}_{-49}$ and $85^{+101}_{-46}$ keV in each, similar to what found for both types of BATSE bursts, which manifests the traditional short-hard/long-soft scheme may not be tenable for the certain energy window of a detector. In statistics, we also investigate the consistency of distinct methods for the $E_{\text p}$ estimates and find that Bayesian approach and BAND function can always give consistent evaluations. In contrast, the frequently-used cut-off power-law model matches two other methods for lower $E_{\text p}$ and will overestimate the $E_{\text p}$ more than 70\% as $E_{\text p}>$100 keV. Peak energies of X-ray flashes, X-ray rich bursts and classical gamma-ray bursts could have an evolutionary consequence from thermal-dominated to non-thermal-dominated radiation mechanisms. Finally, we find that the $E_{\text p}$ and the observed fluence ($S_{\gamma}$) in the observer frame are correlated as $E_p\simeq [S_{\gamma}/(10^{-5} erg\ cm^{-2})]^{0.28}\times 117.5^{+44.7}_{-32.4}$ keV proposed to be an useful indicator of GRB peak energies.

\end{abstract}

\keywords{gamma-ray burst: general ---methods: data analysis---radiation mechanisms: general}


\section{Introduction} \label{sce:intro}
Gamma-Ray Bursts (GRBs) as the strongest and brightest explosions in the universe were first seen in 1967 by the U.S. Vela satellite. It has led more and more people to pursue the nature of its formation, structure and evolution \citep[e.g.,][]{Zhang-04,Piran-04,Meszaros-06,Nakar-07} since the first GRB phenomenon was reported by \cite{Klebesadel+73}. The duration parameter ($T_{90}$) of prompt $\gamma$-rays is defined as the time over which a burst emits a middle 90\% of its total measured photon counts by satellites. In terms of the $T_{90}$, it is found that GRBs can be usually divided into long soft and short hard classes whose boundary is around 2 seconds. The bimodal category of $T_{90}$ is supported by not only the Burst and Transient Source Experiment (BATSE: 20 keV-10 MeV), the \textit{Swift} Burst Alert Telescope (BAT: 15-150 KeV) but also the \textit{Fermi} Gamma-ray Burst Monitor (GBM: 8 keV-40 MeV) data \citep[e.g.,][]{Kouveliotou+93,Zhang+08,Zhang+16,Zitouni-15,Tarnopolski+17,Zitouni-18}, although the BAT detector prefers softer $\gamma$-rays than the other two monitors. However, some authors insisted the number of subgroups in GRBs to be three \citep{Chattopadhyay+07,Horvath+16} or even five \citep{Chattopadhyay+18,Toth+19}. Very recent investigations of the skewed distributions in the plane of hardness ratio versus $T_{90}$ confirm these additional components to be likely artificial \citep{Tarnopolski+19a,Tarnopolski+19b}. Therefore, the classification of GRBs according to $T_{90}$ is still controversial and urgent particularly in view of the spectral properties \citep{Zhang-09}. Under this circumstance, how does the $T_{90}$ with good spectrum measurement distribute is a meaningful issue required to be solved.

An important parameter of $\nu F_{\nu}$ spectrum of the prompt $\gamma-$rays is the peak energy $E_{\text p}$, which presents general spectral properties of GRBs. In the past decades, many authors had studied its statistical distributions and found the typical $E_{\text p}$ is distributed over a broad energy range from a few keV to MeV \citep[e.g.,][]{Preece+00,Sakamoto+04}. Usually, bright GRBs would have larger energy outputs peaking around hundreds of keV to sub-MeV \citep{Preece+16}, while it is much similar to the $E_{\text p}$ distribution of short faint GRBs detected by Konus-wind \citep{Svinkin+16}. Some less luminous GRBs are named as X-ray rich GRBs (XRBs) with $E_{\text p}<$100 keV typically. There are more softer $\gamma$-ray events called as X-ray flashes (XRFs) with spectra peaking below $\sim$30 keV \citep{Barraud+05,Sakamoto+05}. \cite{Katsukura+20} showed that XRBs and XRFs are not the isolated populations from bright GRBs but a phenomenal extension on basis of \textit{Swift}'s multi-wavelength observations. The photosphere model is usually applied to interpret $E_{\text p}$ depending on the
photon temperature of an fireball as it cools adiabatically \citep{Ryde-04,Beloborodov+13}. While the dissipative photosphere model predicts $E_{\text p}$ resulting from the electron temperature \cite{Giannios+12}. Another important mechanism of $E_{\text p}$ formation is related with either internal shock \citep[e.g.,][]{Bosnjak+14,Yu+15} or Internal-collision-induced Magnetic Reconnection and Turbulence (ICMART, \citealt{ZhangB-11}). The latter is controlled by the synchrotron radiation of relativistic electrons with an energy distribution of $N(e)\propto\gamma_{e}^{-p}$, in which $p$ is power law index and $\gamma_e(>\gamma_{min})$ denotes the electronic Lorentz factor. Regarding the radiation processes generating lower $E_{\text p}$ of XRFs, several theories such as high redshift GRB model, off-axis jet model, and sideways expanding opening-angle model have been summarized in \cite{Katsukura+20}.

\begin{figure}
    \centering
    \includegraphics[width=0.8\textwidth]{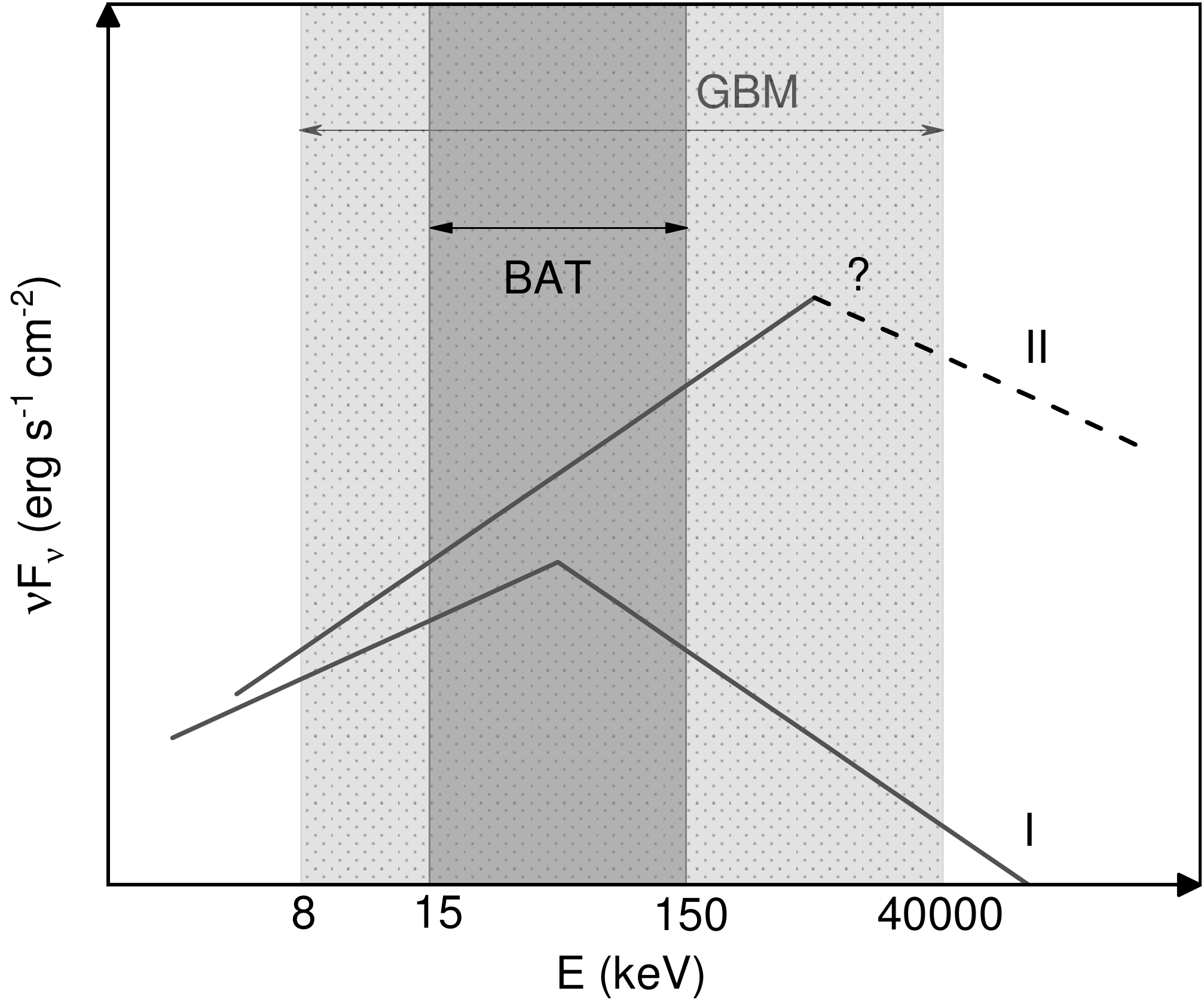}
    \caption{Sketch of two representative types of GRB spectra located in \textit{Swift}/BAT (dark gray) and Fermi/GBM (light gray) energy windows.}
    \label{fig1:Epcases}
\end{figure}

On the other hand, the observed $E_{\text p}$ distribution is easily biased by the energy window of different $\gamma$-ray detectors if only the energy coverage is not wide enough. For example, \cite{Zhang+18} studied the $E_{\text p}$ distributions and energy correlations of $E_{\text p}$ vs. peak luminosity ($L_{\text p}$), $E_{\text p}$ vs. isotropic energy ($E_{iso}$) and $E_{\text p}$ vs. jet-corrected energy ($E_{\gamma}$) for large samples including 31 short and 252 long GRBs with measured $E_{\text p}$, of which 160 and 105 bursts are taken from BAT and GBM catalogs between December 2004 and November 2017, respectively. They found the mean values to be $\langle E_{\text p}\rangle=98\pm2$ and $120^{+38}_{-28}$ keV for long and short GRBs, correspondingly, which is in agreement with the $E_{\text p}$ distribution of 80 \textit{Swift} long GRBs gained by \cite{Katsukura+20} recently. Due to the broader energy bands of BATSE and GBM, extensive multi-wavelength observations of the two satellites give the similar $E_{\text p}$ distributions with a mean value of $\sim$200 keV \citep{Preece+00,Goldstein+13,Gruber+14,Preece+16}. By contrast, the peak energies of BATSE/GBM GRBs are on average two times larger than those of \textit{Swift}/BAT bursts owing to the diverse energy bands of detectors. This leads to a significant fraction of XRFs and XRBs with lower $E_{\text p}$ trigged by \textit{Swift} satellite easily \citep{Butler+07,Katsukura+20}. As shown in Fig.~\ref{fig1:Epcases}, \textit{Swift}/BAT with narrower energy window can only detect lower $E_{\text p}$ of case I, if the $E_{\text p}$ of a burst is located outside of the BAT energy range of as displayed for case II, how can we know where the GRB spectrum would peak? Of course, if only the case II occurs in a wider energy window of detectors like Fermi/GBM as illustrated in Fig.~\ref{fig1:Epcases}, the $E_{\text p}$ can be jointly measured with the aid of other satellites \citep{Zhang+18,Katsukura+20}. Besides, one may apply the statistically Bayesian method \citep{Butler+07} or the empirically Comptonized model \citep[e.g.,][]{Sakamoto+08,Lien+16} to infer the $E_{\text p}$ in both cases I and II. However, whether these inferred peak energies are consistent with each other or not is a fundamental and crucial problem. Furthermore, the $E_{\text p}$ diversity among different energy channels also motivates us to explore the possible emission mechanisms for its physical origins.

In \S\ref{sec:sample} we present data preparation and methodology used in the measurement of $E_{\text p}$. In \S\ref{sec:testing T90} we show the $T_{90}$ distribution of GRBs with well-constrained $E_{\text p}$. Some results related with the empirical $E_{\text p}$ are given in to disclose the underlying import of the $E_{\text p}$ observations on the dominant radiation mechanisms. Conclusions and discussions are given in \S\ref{sec:summary}.-\textit{--to be rewritten}

\section{Methods and Data}
\label{sec:sample}
There are three empirical methods that are popularly used to estimate the $E_{\text p}$ as follows. In general, most fluence-averaged spectra of GRBs can be successfully fitted by the following BAND function \citep{Band+93}
\begin{equation}
\label{eqn:band}
N_{\text{E,BAND}}(E) = A_o\left\{
\begin{array}{ll}
	\left(\frac{E}{100\text{ keV}}\right)^\alpha \exp\left[-\frac{(\alpha+2)E}{E_\text{p,o}}\right]\\ \text{ for } E<E_\text{break}, \\
	\left(\frac{E}{100\text{ keV}}\right)^\beta \exp\left(\beta-\alpha\right) \left(\frac{E_\text{break}}{100\text{ keV}}\right)^{\alpha-\beta} \\ \text{ for } E\geq E_\text{break},
\end{array}
\right.
\end{equation}
where $A_o$ is the normalization factor at 100 keV in units of photons s$^{-1}$cm$^{-2}$keV$^{-1}$, both $\alpha$ and $\beta$ are respectively low and high power-law indices in photon number. Throughout the paper, the $E_{p,o}$ given by the BAND function fit is called the observed peak energy in $E^2N_E$ or $\nu F_\nu$ spectrum and $E_\text{break}=E_\text{p,o}(\alpha-\beta)/(\alpha+2)$ denotes the break energy in the region where two power-law segments connect. To guarantee to yield the peak energy, $\alpha>-2$ and $\beta<-2$ are typically required. The Eq. (\ref{eqn:band}) is valid for the cases I and II only when the GRB spectrum peaks within the energy window of a detector. Observationally, only about one-third of \textit{Swift} GRBs within the BAT window can be directly fitted with this method. The treatment in the case II for \textit{Swift }bursts is to combine the data with other instruments with broader energy region in order to infer the expected $E_{p,o}$ jointly.

The second way of estimating $E_{\text p}$ is to use the Cut-off power-law (CPL) or Comptonized model \citep{Sakamoto+08} in that Eq. (\ref{eqn:band}) will degenerate into the subset form in the limit of $\beta\rightarrow\infty$ as
\begin{equation}
\label{eqn:cpl}
N_{E,CPL}(E) =A_c\left(\frac{E}{100\text{ keV}}\right)^{\alpha^{CPL}} \exp\left[-\frac{(\alpha^{CPL}+2)E}{E_\text{p,c}}\right],\\
\end{equation}
in which $A_c$ is the normalization factor at 100~keV in units of photons~s$^{-1}$~cm$^{-2}$~keV$^{-1}$, $\alpha^{CPL}$ and $E_{\text {p,c}}$ are spectral index and peak energy, respectively. The third method is the popular and unbiased Bayesian approach \citep{Butler+07} employed to estimate the peak energy $E_{\text {p,b}}$ in the posterior distribution of probability as
\begin{equation}
\label{eqn:bayesian}
P(lnE_\text{p,b})\propto exp\left\{-0.5\left[ln\left(\frac{E_\text{p,b}}{300 keV}\right)\right]^2/\sigma^{2}_{lE_{\text p}}\right\}, \\
\end{equation}
where the zero probability has been assumed for $E_\text{p,b}>10^4$ keV and $E_\text{p,b}<1$ keV. At the same time, two parameters $\alpha$=-1.1 and $\beta$=-2.3 gotten from the BATSE data catalog in Eq. (\ref{eqn:band}) have been set for two prior exponential probability distributions \citep{Butler+07}. It is noticeable that the CPL and the Bayesian approaches are initially proposed to estimate peak energies for the case II but available for the case I as well. We emphasize that Eqs. (2) and (3) are two separate ways to infer the $E_{\text p}$ no matter where the GRB spectra peak.

For the purpose of understanding physical properties of prompt $\gamma$ emissions, the observed GRB spectra are usually ``fitted'' to any one of the chosen models in Eqs.(\ref{eqn:band})-(\ref{eqn:bayesian}). Table \ref{table1} gives the comparisons between these empirical peak energies. To investigate the consistency of different $E_{\text p}$ determinations, we select \textit{Swift}/BAT bursts with well-constrained $E_{\text p}$ detected from December 2004 and October 2018. Our sample selection criteria are defined as follows: (1) use \textit{Swift}/BAT GRBs only; (2) select GRB spectra fitted with BAND function successfully, regardless of $E_{\text p}$ inside or outside of energy window of \textit{Swift}/BAT, and (3) choose those GRBs with measured $E_{\text {p,o}}$ at a higher confidence level of $S/N>3$. Thus, out of 1256 \textit{Swift}/BAT bursts, 283 including 15 short and 268 long GRBs with both $E_{\text {p,o}}$ and $E_{\text {p,b}}$ match the above standards and are directly taken from Nat Butler's Swift BAT+XRT(+optical) repository\footnote{http://butler.lab.asu.edu/Swift/index.html}. Totally, 266 consisting of 15 short and 251 long GRBs with known $E_{\text {p,c}}$ are obtained from the \textit{Swift}/BAT Gamma-Ray Burst Catalog \footnote{https://swift.gsfc.nasa.gov/results/batgrbcat/}. In addition, we also collect the $T_{90}$ values for 279 GRBs from the official \textit{Swift} GRB table\footnote{https://swift.gsfc.nasa.gov/archive/} and 4 GRBs (090621A, 100514A, 130306A and 130807A) from the Gamma-ray Coordinates Network\footnote{https://gcn.gsfc.nasa.gov/}, and the redshifts of 92 GRBs from JG's homepage\footnote{http://www.mpe.mpg.de/~jcg/grbgen.html}. It is noteworthy that in our sample there are 37 GRBs with $E_{\text {p,o}}>$ 150 keV that is already beyond the \textit{Swift}/BAT energy window.
\begin{table}
	\centering
	\caption{Summary of different peak energies in this work}
	\begin{tabular}{|c|c|c|c|c|}
 	\hline
   		Peak energy & Definition & Property & Comments & Reference  \\
	\hline
       $E_{\text {p,o}}$ & by Eq.(\ref{eqn:band}) & observed & measured+empirical & \cite{Band+93}  \\
       $E_{\text {p,c}}$ & by Eq.(\ref{eqn:cpl}) & Comptonized & inferred+empirical  & \cite{Sakamoto+08}  \\
       $E_{\text {p,b}}$ & by Eq.(\ref{eqn:bayesian}) & Bayesian & inferred+empirical   & \cite{Butler+07}  \\
       \hline
       $E_{\text {p,BB}}$ & by Eq.(\ref{eqn:bb}) & theorized & simulated+physical & \cite{Rybicki-79}  \\
       $E_{\text {p,syn}}$ & by Eq.(\ref{eqn:sync}) & theorized & simulated+physical & \cite{Rybicki-79}  \\
        \hline
       $E_{\text {p,i}}$ & $E_{\text {p,o}}$(1+z) & intrinsic & measured+empirical  & \cite{Dodelson-03}  \\
    \hline
   	\end{tabular}
      \label{table1}
\end{table}

\section{Durations and Classifications}
\label{sec:testing T90}
\begin{figure}
    \centering
    \includegraphics[width=0.8\textwidth]{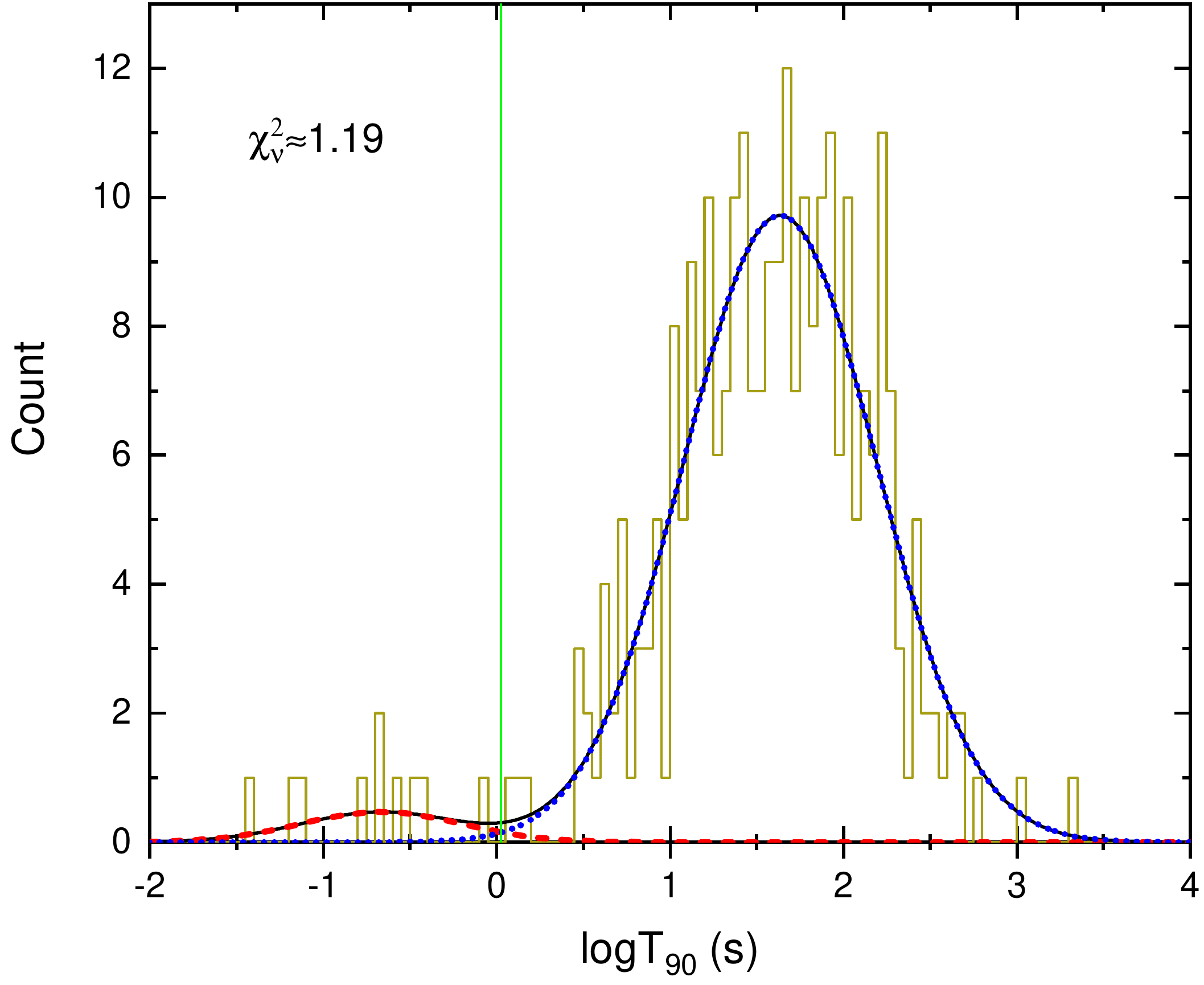}
    \caption{The $T_{90}$ distribution of 283 \textit{Swift} GRBs with well measured $E_{\text {p,o}}$. A double-gauss model fitted to data is marked by the solid black line with a reduced Chi-square $\chi^2_{\nu}\simeq1.19$ and a dividing line near 1.06 s. Short and long GRBs are symbolized by dashed and dotted lines, respectively.}
    \label{fig2:t90dis}
\end{figure}

In this section, we display the temporal and spectral results of both short and long GRBs with well-confirmed $T_{90}$ and $E_{\text p}$ simultaneously. In the past dozen years, several authors studied the $T_{90}$ distribution of \textit{Swift} GRBs and found that bimodal rather than triple GRB groups are preferred \citep[e.g.,][]{Zhang+08,Zhang+16,Zitouni-15,Zitouni-18}. Fig. \ref{fig2:t90dis} shows the $T_{90}$ distribution of 283 \textit{Swift} bursts with confident $E_{\text {p,o}}$ at $\gtrsim 3\sigma$ level. The two-gaussian fit to data demonstrates that the duration distribution peaks at $0.21^{+0.31}_{-0.12}$ s with a spread of 0.93 dex for short bursts and at $42.66^{+2.00}_{-1.92}$ s with a spread of 1.11 dex for long GRBs. The best fit returns a good reduced Chi-square of $\chi^2_{\nu}\simeq1.19$ indicating two classes are evidently reconfirmed and separated at $T_{90}\simeq$1.06 s other than 2 s shown by CGRO/BATSE data \cite{Kouveliotou+93} or \textit{Swift}/BAT normal observations by \citep{Zhang+08,Zhang+16}. However, the new dividing line at $T_{90}$=1.06 s is very close to 1.27 s of GBM GRBs with precise $T_{90}$ measurements at a S/N larger than 3 (see Fig. 1a in \citealt{Gruber+14}). It is noticeable that the ratio of short to long GRBs is about $\lesssim$1/10 for \textit{Swift} against $\sim$1/4 for Fermi or BATSE mission, which is deeply suffered from the threshold effect \citep{Zhang+16}. Additionally, we point out that the durations of \textit{Swift}/BAT long bursts have relatively wider distribution and are systematically longer than those of GBM/BATSE long GRBs. This is attributed to the fact that BAT is more sensitive to long softer GRBs \citep{Gehrels+04}, especially those short GRBs with extended emission that would easily confuse the classification in terms of $T_{90}$ \citep{Zhang+16}.

\section{Properties of Empirical Peak energies}
\label{sec:empirical Ep}
Here, consistency of the above different inferences of peak energy is to be checked in statistics. Taking into account the $E_p$ diversity, we subsequently reexamine the peak energy versus the fluence in the observer frame.
\subsection{Comparing Peak Energies between Different Models}
\label{sec:Ep-compare}
\begin{figure}
    \centering
    \includegraphics[width=0.8\textwidth]{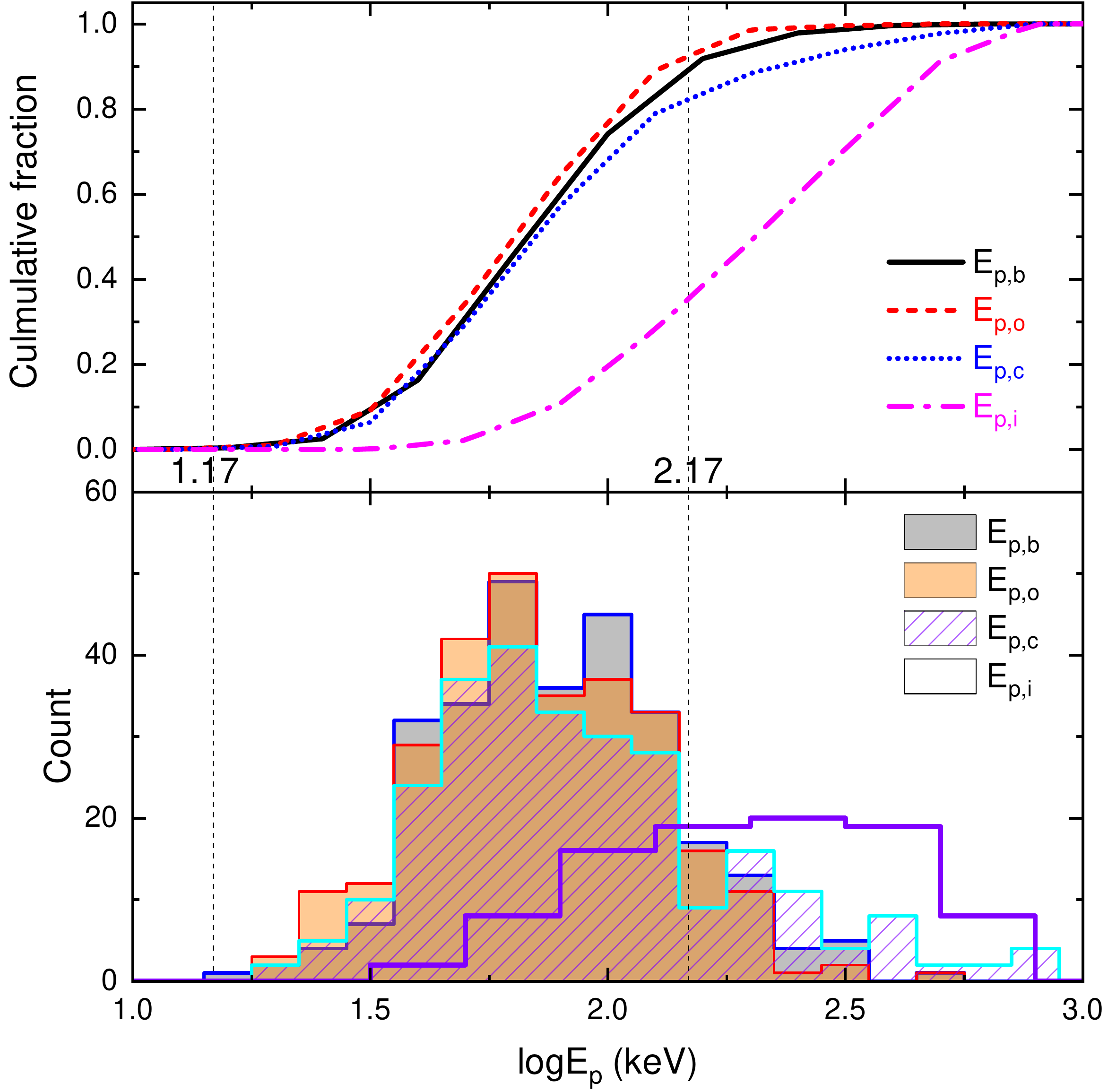}
    \caption{Comparisons between three lab-frame peak energies inferred from diverse methods along with the $E_{p,i}$ in the rest frame. Two vertical dashed-lines represent the energy ranges of \textit{Swift}/BAT in logarithmic scale.  }
    \label{fig3:allGRBEpdis}
\end{figure}

We now move to investigate the similarities and differences of peak energies given by different methods illustrated in \S\ref{sec:sample}. Fig. \ref{fig3:allGRBEpdis} is plotted for the cumulative fraction distributions of all selected GRBs in upper panel and their histograms in lower panel for three kinds of $E_{\text p}$ in both the observer and the rest frames, from which we are apt to see the obvious difference between the intrinsic $E_{\text {p,i}}=(1+z)E_{\text {p,o}}$ and the observed $E_{\text {p,o}}$ distributions. In fact, a Kolmogorov-Smirnov (K-S) test to the $E_{\text {p,o}}$ and $E_{\text {p,i}}$ distributions gives $D=0.66$ with a \textit{p}-value of $9.48\times10^{-25}$ showing that they are not drawn from the same parent distribution, of which the median of the $E_{\text {p,i}}$ is about three time larger than that of the $E_{\text {p,o}}$. Although $E_{\text {p,o}}$, $E_{\text {p,b}}$ and $E_{\text {p,c}}$ seem to be identically distributed, an Anderson-Darling (A-D) test demonstrates in the first set of Table \ref{adtest} that $E_{\text {p,o}}$ and $E_{\text {p,c}}$ are differently distributed, while the other two pairs of peak energies share the same distributions. The observer-frame mean values of peak energies are correspondingly $E_{\text {p,o}}\simeq85.1^{+62.8}_{-36.1}$ keV, $E_{\text {p,b}}\simeq79.4^{+58.6}_{-39.4}$ keV and $E_{\text {p,c}}\simeq99.9^{+108.9}_{-52.2}$ keV, of which all are lower than the rest-frame peak energy of $E_{p,i}\simeq245.5^{+244.3}_{-122.5}$ keV and the scatter of $E_{\text {p,c}}$ measurement is comparably larger than that of either $E_{\text {p,o}}$ or $E_{\text {p,b}}$ in the observer frame. Note that we have used Fermi/GBM data to constrain the $E_{\text {p,o}}$ of a small fraction ($\sim 13\%$) of \textit{Swift} bursts as their peak energies exceed the BAT upper limit of 150 keV.

\begin{figure}
    \centering
    \includegraphics[width=0.8\textwidth]{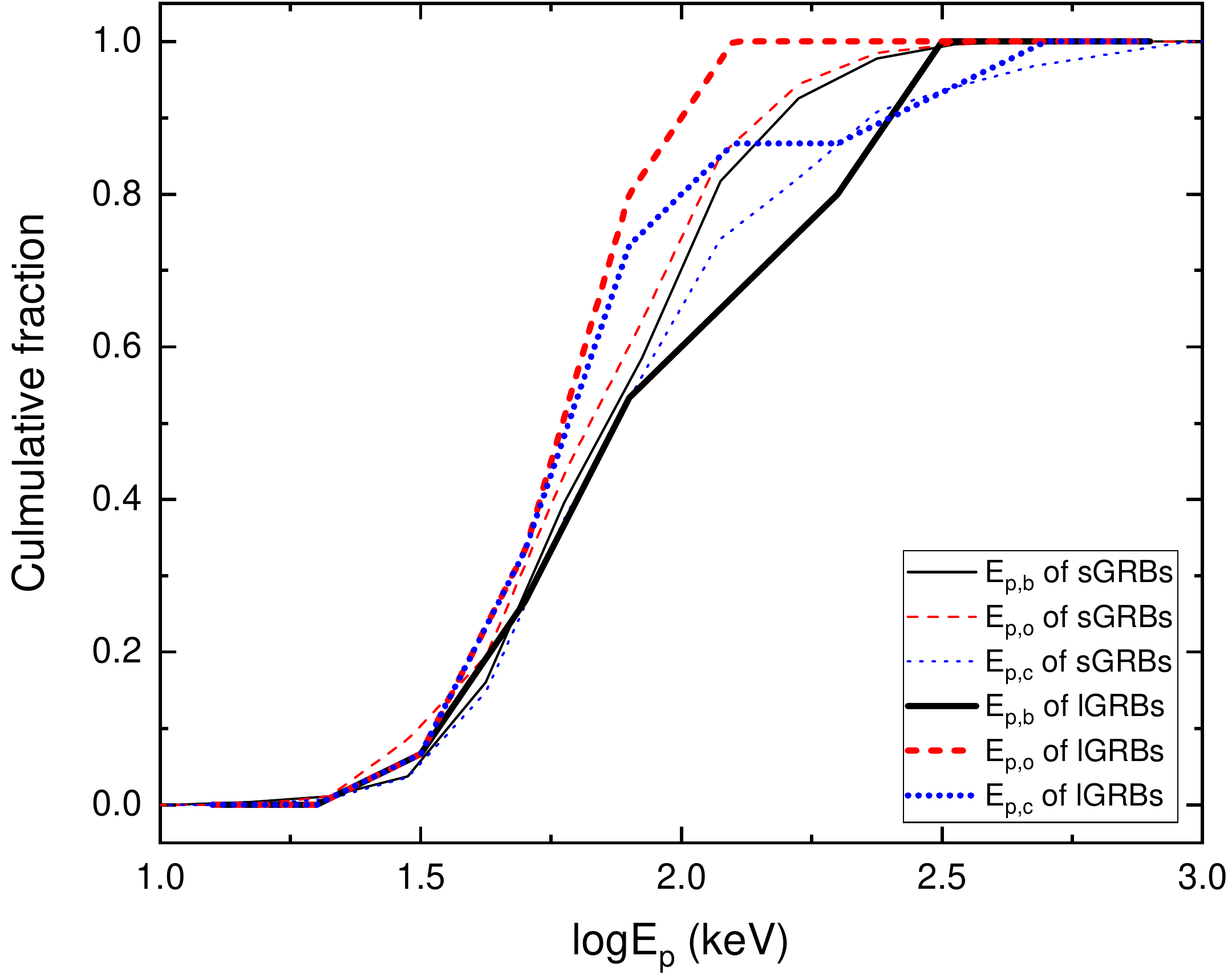}
    \caption{Cumulative fractions of different $E_{p}$ for \textit{Swift}/BAT short (thin lines) and long (thick lines) GRBs, respectively. }
    \label{fig4:longshortGEBEpdis}
\end{figure}

The $E_{\text {p,o}}$ generally evolves with time because of the spectral evolution (e.g., \citealt{ZhangB-12}). In particular, most long GRBs with higher redshift exhibit a trend of hard-to-soft spectral evolution \citep{Norris+00,Daigne+03}, while short GRBs with lower redshift show no significantly spectral evolutions \citep{Norris+06,Zhang+06}. It is however not clear whether there is a dependence of the $E_{\text p}$ estimates on redshift for both short and long bursts. We compare the three observer-frame $E_{\text p}$ in Fig. \ref{fig4:longshortGEBEpdis} where peak energies of short bursts are superficially different in distribution from those of long bursts. Surprisingly, the second set of Table \ref{adtest} from the A-D tests demonstrates that the estimates of $E_{\text {p,o}}$, $E_{\text {p,b}}$ and $E_{\text {p,c}}$ are consistent with each other for short GRBs, and similar to long bursts whose $E_{\text p}$ distributions are statistically same except for the inconformity between $E_{\text {p,o}}$ and $E_{\text {p,c}}$. One can logically infer that the distributional difference of $E_{\text {p,o}}$ from $E_{\text {p,c}}$ in the complete $E_{\text p}$ sample is sightly caused by long GRBs and the $E_{\text p}$ estimates based on different models are not connected with redshifts or durations in evidence. Interestingly, our highly confident peak energy samples return the consistent mean values of $E_{\text {p,o}}$ to be $\sim87^{+112}_{-49}$ keV and $\sim85^{+101}_{-46}$ keV for short and long GRBs, respectively, which greatly deviates from our traditional comprehension of short-hard versus long-soft bursts. The possible reason is that most \textit{Swift} short GRBs are accompanied by the EE components softening GRB energy spectra, while they are easily detected thanks to \textit{Swift}'s instrumental design. Noticeably, \cite{Ghirlanda+04} also found the averaged $E_{\text p}$ values of short and long bursts to be statistically same even for the BATSE detector with a broader energy coverage. Hence one can conclude that the $E_{\text p}$ is not an optimized parameter indicating the hardness of GRB spectra as two classes of GRBs could be essentially identical on aspect of the radiation mechanism. More interestingly, this is very similar to previous conclusions based on the relations of hardness ratio (HR) with $T_{90}$ for \textit{Swift}/BAT short and long GRBs with fluences better-fitted by a spectral model \citep{Sakamoto+11,Zhang+16,Lien+16}.

\begin{figure*}
    \centering
    \includegraphics[width=1.0\textwidth]{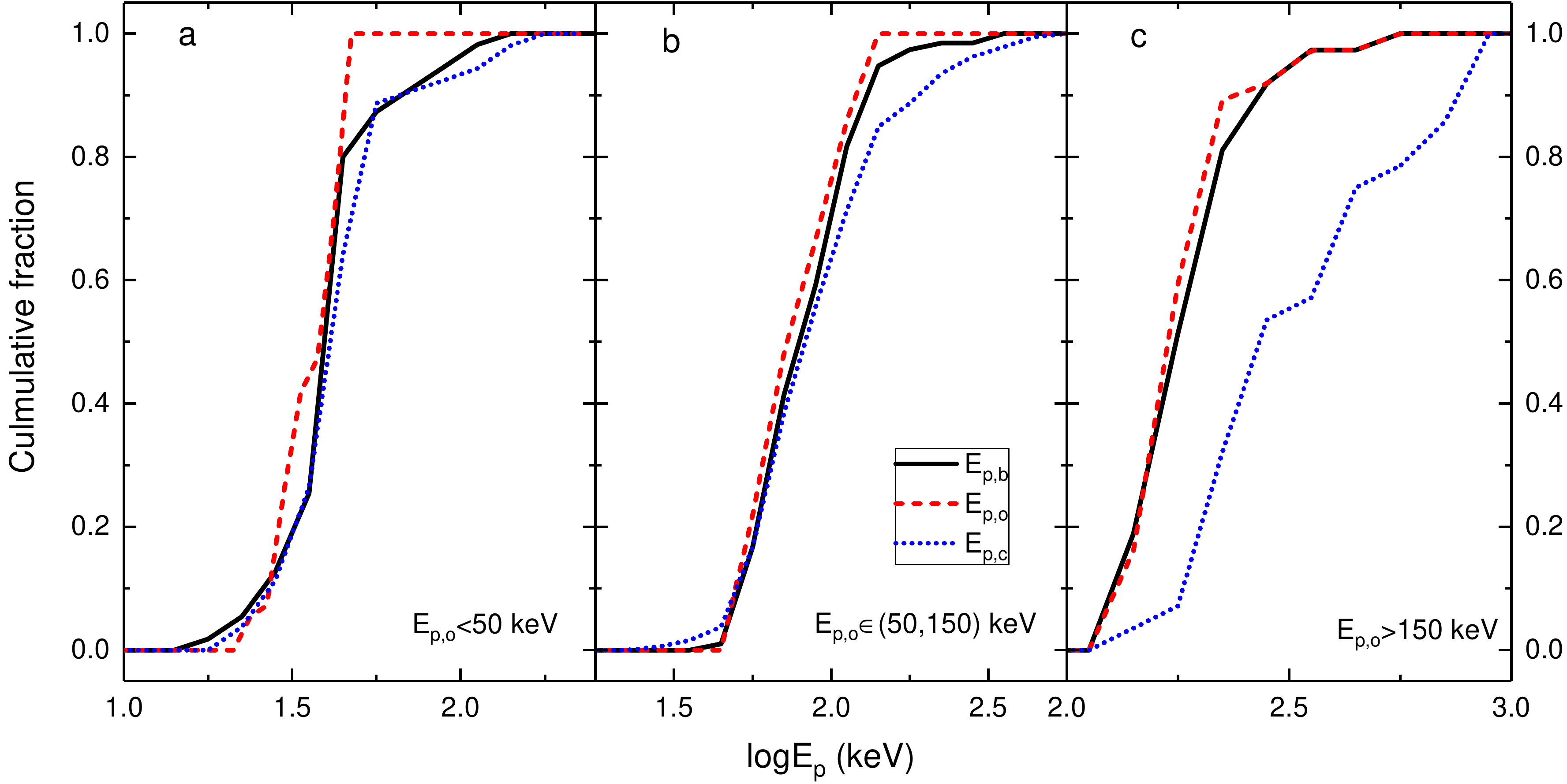}
    \caption{Cumulative fractions of different $E_{p}$ for \textit{Swift}/BAT GRBs in low (LE in panel a), middle (ME in panel b) and high (HE in pane c) energy bands, respectively.}
    \label{fig5:differentEpdis}
\end{figure*}

Furthermore, we study the distributional consistency of three kinds of $E_{\text {p}}$ in Fig. \ref{fig5:differentEpdis}, from which the burst numbers in each energy channel are respectively 55 for low energy (LE) band below 50 keV, 191 for middle energy (ME) band between 50 and 150 keV, and 37 for high energy (HE) band above 150 keV. Notably, the diverse energy bands roughly correspond to the energy scopes of XRFs, XRBs and classical-GRBs (C-GRBs) defined by \citet{Katsukura+20}. The A-D tests have also been done and returned the results in the third set of Table \ref{adtest}, in which we find that $E_{\text {p,b}}$ and $E_{\text {p,c}}$ share with the same distribution ($AD$=0.91 and $p$=0.41) in the LE wavelength, while $E_{\text {p,b}}$ and $E_{\text {p,o}}$ are identically distributed in both ME and HE wavelengths due to smaller $AD$ statistics and larger $p$-values. Moreover, if considering Figs. 3-5 as a whole, we notice that the Bayesian approach in Eq. (\ref{eqn:bayesian}) can always return very similar $E_{\text {p,b}}$ to those $E_{\text {p,o}}$ fitted by the BAND function in Eq. (\ref{eqn:band}). On the contrary, the $E_{\text {p,c}}$ estimated by the CPL model of Eq. (\ref{eqn:cpl}) can not ideally match the observed peak energies given by Eq. (\ref{eqn:band}) for all classes of GRBs but the short ones. It needs to be emphasized that $E_{\text {p,c}}$ compared with either $E_{\text {p,b}}$ or $E_{\text {p,o}}$  will be overestimated when the peak energies are larger than $\sim100$ keV.

\begin{longrotatetable}
\begin{deluxetable*}{ccccccc}
\tablecaption{Parameters of A-D test between different $E_{\text p}$ distributions for a significance level of $\alpha=0.01$\label{adtest}}
\tablewidth{700pt}
\tabletypesize{\normalsize}
\tablehead{
\colhead{GRB type} & \colhead{$E_{\text p}$ pair} &
\colhead{Sample size} & \colhead{$AD-{stat}$} &
\colhead{ $p$-value} & \colhead{$AD-{crit}$} &
\colhead{Null hypothesis}
}
\startdata
	Complete & $E_{\text {p,b}}$:$E_{\text{p,o}}$ &283:283& 1.614 & 0.151 & 3.8705& accepted  \\
    Complete & $E_{\text {p,o}}$:$E_{\text {p,c}}$ &283:266& 5.477 & 0.002 & 3.8702& rejected  \\
    Complete & $E_{\text {p,b}}$:$E_{\text {p,c}}$ &283:266& 2.655 & 0.041 & 3.8702& accepted  \\
	\hline
	Short & $E_{\text {p,b}}$:$E_{\text {p,o}}$ &15:15& 1.264 & 0.239 & 3.7296& accepted  \\
    Short & $E_{\text {p,o}}$:$E_{\text {p,c}}$ &15:15& 0.384 & 0.896 & 3.7296& accepted  \\
    Short & $E_{\text {p,b}}$:$E_{\text {p,c}}$ &15:15& 0.596 & 0.673 & 3.7296& accepted  \\
    Long & $E_{\text {p,b}}$:$E_{\text {p,o}}$ &268:268& 1.290 & 0.235 & 3.8700& accepted  \\
    Long & $E_{\text {p,o}}$:$E_{\text {p,c}}$ &268:251& 5.416 & 0.002 & 3.8697& rejected  \\
    Long & $E_{\text {p,b}}$:$E_{\text {p,c}}$ &268:251& 3.192 & 0.022 & 3.8697& accepted  \\
    \hline
    LE& $E_{\text {p,b}}$:$E_{\text {p,o}}$ &55:55& 5.279 & 0.002 & 3.8377&rejected \\
    LE& $E_{\text {p,o}}$:$E_{\text {p,c}}$ &55:53& 8.454 & 5.7E-5 & 3.8370&rejected  \\
    LE& $E_{\text {p,b}}$:$E_{\text {p,c}}$ &55:53& 0.913 & 0.407 & 3.8370& accepted  \\
    ME& $E_{\text {p,b}}$:$E_{\text {p,o}}$ &191:191& 1.769 & 0.123 & 3.8667& accepted  \\
    ME& $E_{\text {p,o}}$:$E_{\text {p,c}}$ &191:185& 8.284 & 7.7E-5 & 3.8665& rejected  \\
    ME& $E_{\text {p,b}}$:$E_{\text {p,c}}$ &191:185& 3.902 & 9.6E-3 & 3.8665& rejected  \\
    HE& $E_{\text {p,b}}$:$E_{\text {p,o}}$ &37:37& 0.599 & 0.656 & 3.8181& accepted  \\
    HE& $E_{\text {p,o}}$:$E_{\text {p,c}}$ &37:\textbf{28}& 10.727 & 7.6E-7 & 3.8097& rejected \\
    HE& $E_{\text {p,b}}$:$E_{\text {p,c}}$ &37:\textbf{28}& 9.808  & 6.5E-6 & 3.8097& rejected  \\
    \enddata
\end{deluxetable*}
\end{longrotatetable}


\subsection{Peak Energy versus Fluence }
\label{sec:Epvsfluence}
Using the complete BATSE 5B Spectral Catalog, \cite{Goldstein+10} found that roughly $65\%$ of the bolometric fluence ($S_{bolo}$) distribution for short GRBs overlaps that for long GRBs with the peak position being an order of magnitude larger than the fluence distribution of short GRBs. Here, the ratio of the observed fluence ($S_{\gamma}$) overlapping of short to long GRBs in our \textit{Swift}/BAT GRB sample is approximately 80\% and their discrepancy is similar. Very recently, \cite{Katsukura+20} studied the spectral properties of 80 \textit{Swift} long bursts (26 XRFs, 41 XRBs and 13 C-GRBs) with well-constrained $\gamma$-ray and X-ray spectral parameters and found that three subclasses can be clearly divided at 30 and 100 keV according to the fluence ratio between channels 1 and 2 of the BAT detector. Simultaneously, they showed a weaker dependence of $E_{\text {p,o}}$ on $S_{\gamma}$ and the XRFs are slightly dimmer than the other two kinds of bursts.

\begin{figure}
    \centering
    \includegraphics[width=0.8\textwidth]{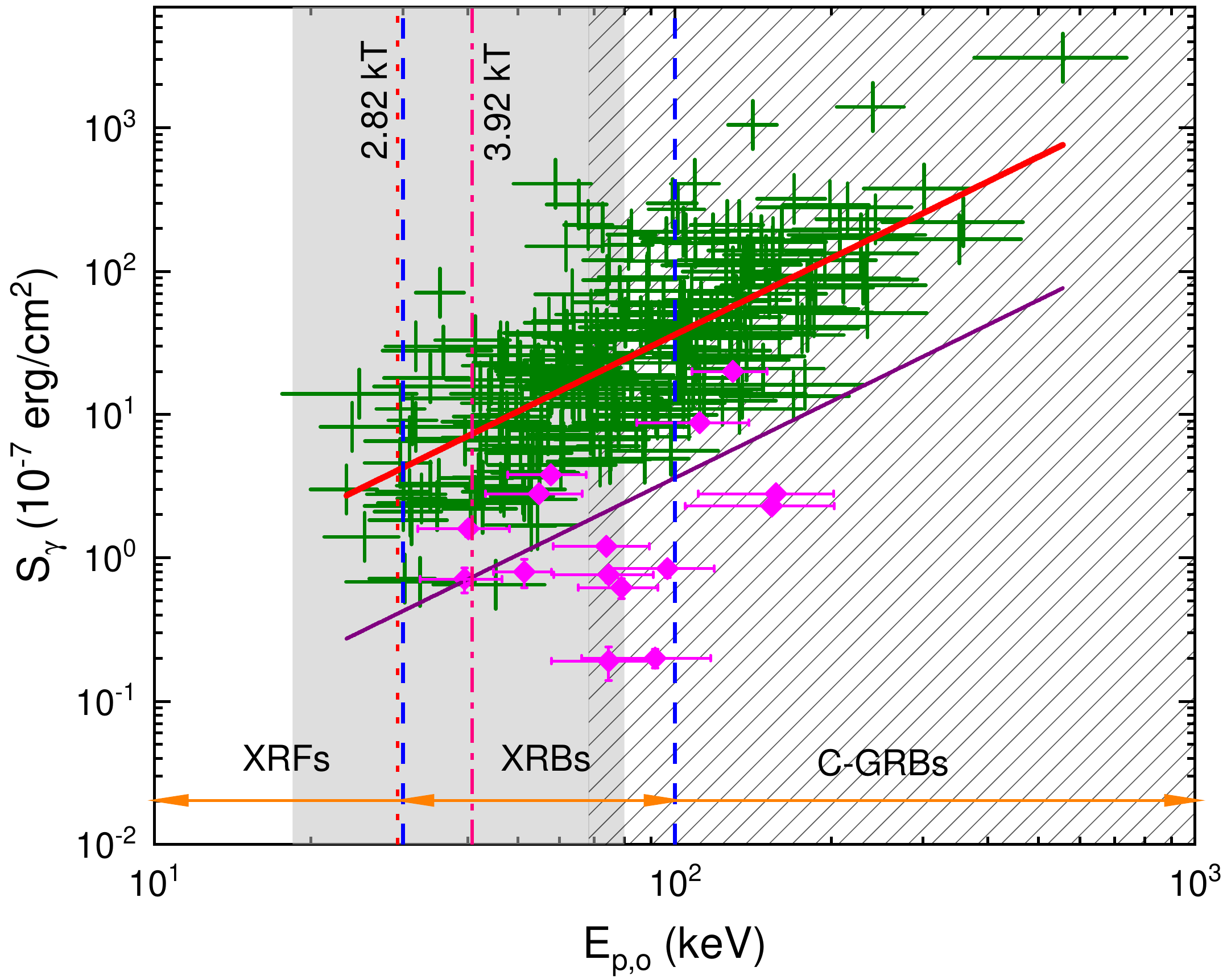}
    \caption{The observed fluence $S_{\gamma}$  is plotted against the observed peak energy $E_{\text {p,o}}$ for 15 short (diamonds) and 268 long (crosses) \textit{Swift}/BAT bursts in our sample. Two vertically dashed lines at 30 and 100 keV are symbolized to divide XRFs, XRBs and C-GRBs. The dotted and the dash-dotted red lines represent two theoretical peak energies of the BB spectra (see text). The upper solid line denotes the best logarithmic fit to the long GRBs. The lower solid line is plotted for the short GRBs with 10\% of fluences of long ones.}
    \label{fig6:Ep-fluence}
\end{figure}

By contrast, our sample including both 268 long and 15 short GRBs with well-determined spectra is largely expanded. The $E_{\text {p,o}}-S_{\gamma}$ relations in the observer frame for the total 283 \textit{Swift} GRBs are displayed in Fig. \ref{fig6:Ep-fluence}, in which one can find that the current relation of $E_{\text {p,o}}$ with $S_{\gamma}$ for \textit{Swift} long GRBs is more tight than some previous ones \citep{Goldstein+10,Zhang+18,Katsukura+20} and can be well fitted by
\begin{equation}
\label{Ep-fluence}
log E_\text{p,o}=(3.47\pm0.09)+(0.28\pm0.02)\times log S_{\gamma},
\end{equation}
with Spearman correlation coefficient R=0.70 and a chance probability $P=6.9\times10^{-31}$. This implies that the Eq. (\ref{Ep-fluence}) can be employed as an indicator to estimate the $E_{\text {p}}$ of a burst without good spectral breaks. However, the $E_{\text {p,o}}-S_{\gamma}$ relation of 15 short GRBs is largely scattered possibly owing to the limited data points. It is noteworthy that our $E_{\text {p,o}}-S_{\gamma}$ relation resembles the relation of $E_{\text {p}}$ versus photon index $\Gamma$ as $log E_{\text {p,o}}\simeq3.26-0.83\Gamma$ proposed by \cite{Sakamoto+09} for \textit{Swift} long GRBs. If combing both of them, one can naturally obtain an expression of $S_{\gamma}\simeq10^{-(2.96\Gamma+0.76)}$ in unit of erg/cm$^{2}$. To analyze the $E_{\text {p,o}}$ diversity as well as its possible origins, we draw two boundaries at $E_{\text {p,o}}=$30 and 100 keV in Fig. \ref{fig6:Ep-fluence} to classify our sample of 283 \textit{Swift }GRBs into three subgroups, that is 6 XRFs, 176 XRBs and 101 C-GRBs. By comparing with \cite{Katsukura+20} in the plane of $S_{\gamma}$ against $E_{\text {p,o}}$, we find that our percentages of XRFs, XRBs and C-GRBs are respectively 2\%, 62\% and 36\% that visibly differ from \cite{Katsukura+20} where their vast majority of \textit{Swift }bursts hold much lower $E_{\text {p,o}}$. However, our ratios of different kinds of bursts are very close to those of the raw \textit{Swift }data. 

\section{Implications for radiation mechanisms}
\label{sec:theoryexplain}
We now focus on the analysis of the potential radiation mechanisms leading to the diverse $E_{\text p}$ over a broader energy scope. In physics, the peak energy of GRBs is usually thought to generate from either thermal/quasi-thermal or non-thermal radiations of relativistic electrons in plasma outflows. In case of the thermal $\gamma$-ray emissions from a dissipative photosphere or blackbody (BB), the observed spectrum can be well described by a Planck function below
\begin{equation}
\label{eqn:bb}
N_\text{E,BB}(E) = A_b\left(\frac{E}{1\text{ keV}}\right)^2\left[\exp\left(\frac{E}{kT}\right)-1\right]^{-1},
\end{equation}
where $A_b$ is the normalization factor at 1~keV in units of photons~s$^{-1}$~cm$^{-2}$~keV$^{-1}$, $k$ is Boltzmann constant, $T$ and $kT$ are respectively absolute temperature in Kelvin and thermal energy in unit of keV of the BB. $N_\text{E,BB}(E)$ reaches its maximum value at a certain peak energy of $E_\text{p,BB}$.

Following \cite{Yu+15}, the synchrotron radiation as the most representative non-thermal mechanism of GRBs can be characterized with the following form
\begin{equation}
\label{eqn:sync}
N_\text{E,syn}(E) = A_s\left\{
\begin{array}{ll}
	\left(\frac{E}{100\text{ keV}}\right)^{\alpha^{\prime}} \ \ \ \ \ \ \  \text{ for } E < E_\text{b,1}, \\
	\left(\frac{E_\text{b,1}}{100\text{ keV}}\right)^{\alpha^{\prime}} \left(\frac{E}{E_\text{b,1}}\right)^{\beta^{\prime}} \ \ \text{ for } E_\text{b,1} \leq E < E_\text{b,2}, \\
	\left(\frac{E_\text{b,1}}{100\text{ keV}}\right)^{\alpha^{\prime}} \left(\frac{E_\text{b,2}}{E_\text{b,1}}\right)^{\beta^{\prime}} \left(\frac{E}{E_\text{b,2}}\right)^{\gamma^{\prime}} \ \ \text{ for } E \geq E_\text{b,2},
\end{array}
\right.
\end{equation}
where $A_s$ is the normalization factor at 100~keV in units of photons~s$^{-1}$~cm$^{-2}$~keV$^{-1}$, $\alpha^{\prime}$, $\beta^{\prime}$ and $\gamma^{\prime}$ are three spectral power-law indices of the corresponding parts, $E_\text{b,1}$ and $E_\text{b,2}$ stand for two break energies in units of keV at some characteristic frequencies, namely the cooling frequency $\nu_c$ and the minimum frequency $\nu_m$ of photons emitted from electrons within slow or fast cooling scenario \citep{Sari+98}. In case of the slow cooling ($\nu_m<\nu_c$), we have $E_\text{b,1}=h \nu_m$, $E_\text{b,2}=h \nu_c$, $\alpha^{\prime}=-2/3$, $\beta^{\prime}=-(p+1)/2$ and $\gamma^{\prime}=-p/2-1$. For the fast cooling ($\nu_c<\nu_m$), the critical parameters are set as $E_\text{b,1}=h \nu_c$, $E_\text{b,2}=h \nu_m$, $\alpha^{\prime}=-2/3$, $\beta^{\prime}=-3/2$ and $\gamma^{\prime}=-p/2-1$. For simplicity, the electronic power-law index of $p=2.4$ is assumed. No matter which case we choose, the $E^2N_E$ spectra will be expected to peak at a given energy $E_{\text {p,syn}}$ ranging from $E_\text{b,1}$ to $E_\text{b,2}$ \citep{Yu+15}.

\begin{figure}
    \centering
    \includegraphics[width=0.8\textwidth]{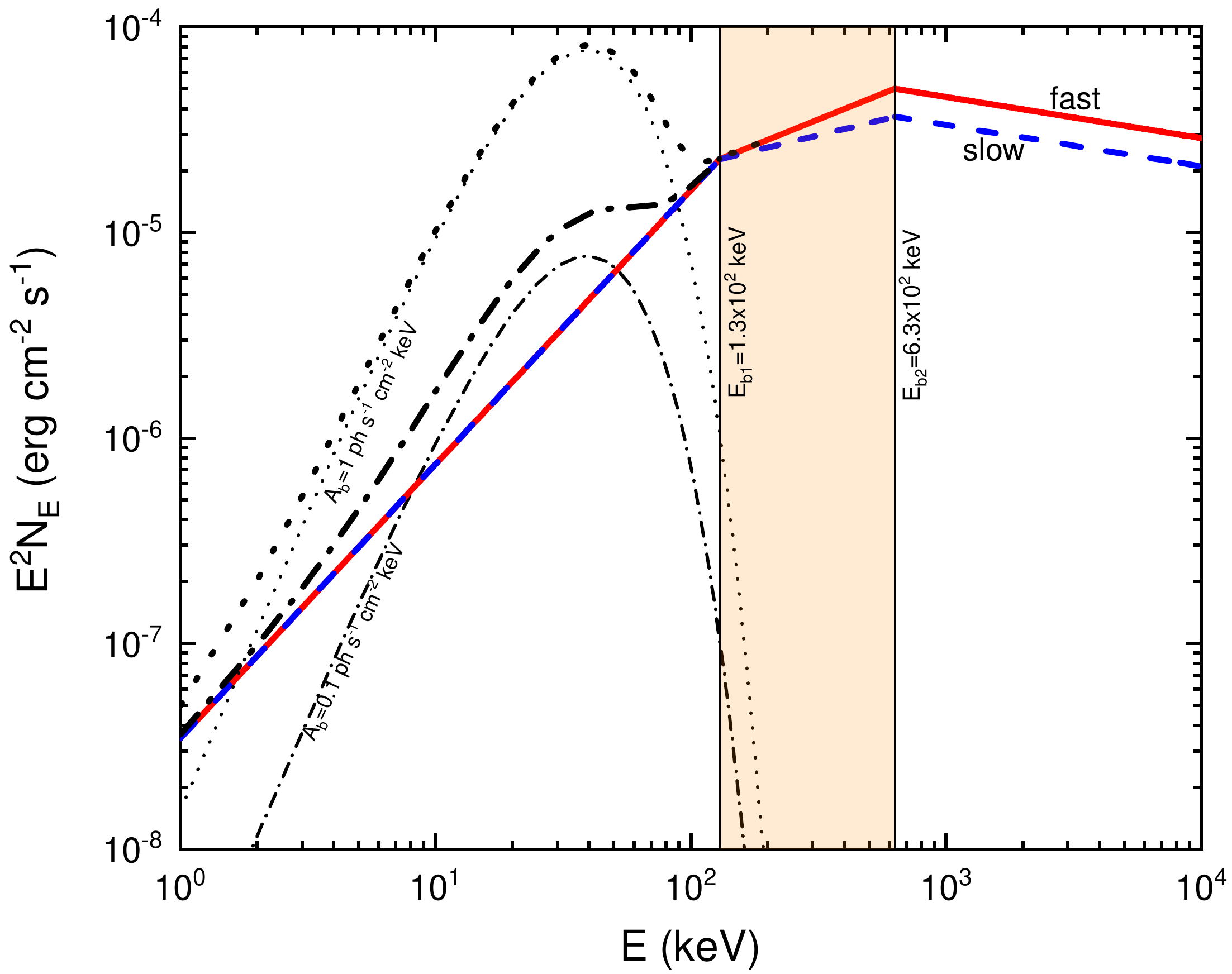}
    \caption{GRB multi-color energy spectra reproduced from BB and Synchrotron components. Thick solid and dashed lines represent Synchrotron emissions with $A_s$=1 photons~s$^{-1}$~cm$^{-2}$~keV$^{-1}$ from the fast and slow electronic cooling, respectively. Thin dash-dotted and dotted lines denote thermal radiations for two typical $A_b$ in text, correspondingly, while their top thick lines are the combinations of them with the Synchrotron segment. Two vertical lines show the representative break energies in the Synchrotron model.}
    \label{fig7:Physical Ep}
\end{figure}
Fig. \ref{fig7:Physical Ep} is plotted with Eqs. (\ref{eqn:bb}) and (\ref{eqn:sync}) to illustrate the potentially physical forming mechanisms of $E_{\text p}$ from the BB and/or the Synchrotron radiations. Here, the average values of $kT\simeq10.4_{-3.7}^{+4.9}$ keV, $E_\text{b,1}\simeq130_{-32}^{+132}$ keV to $E_\text{b,2}\simeq631_{-310}^{+583}$ keV gotten by \cite{Yu+15} have been directly utilized to generate the diverse spectra, of which most GRBs have peak energies observed within $E_\text{b,1}<E_{\text {p,o}}<E_\text{b,2}$. It is easily understood that the standard BB spectrum always peaks at 3.92 $kT$ ($\simeq$40.77 keV) despite of different $A_b$ factors. Here we take $A_b$=0.1 and 1 photons\ s$^{-1}$\ cm$^{-2}$\ keV$^{-1}$ as representative values aligned with the BAT observations. According to the theoretical arguments, the $E_\text{p,BB}$ should has an upper limit or ``death'' line of 2.82 $kT$ for a relativistic outflow \citep{Li+08,ZhangB-12}. Considering the error bars and statistical fluctuations of $E_\text{b,1}$ and $kT$, the predicted $E_{\text p}$ distributions from the BB and the Synchrotron models are to be bridged or overlapped, which is in agreement with the conclusions of Fig.3 drawn by \cite{Yu+15} observationally. At the same time, two 3$\sigma$-confidence ranges of the predicted peak energies from Eqs. (\ref{eqn:bb}) and (\ref{eqn:sync}) are marked with the shaded and hatched regions in Fig. \ref{fig6:Ep-fluence}, respectively. Considering the equivalent GRB numbers within different energy ranges, we infer that around 50\% of \textit{Swift} GRBs are better to be explained by the BB for most XRFs plus XRBs or the Synchrotron radiation mechanisms for most C-GRBs, correspondingly. Instead, only a small fraction of XRBs need to be explained by the combination of the thermal and non-thermal components. It is naturally concluded that there is an evolutionary consequence of $E_{\text {p}}$ from thermal-dominated to non-thermal-dominated emissions at least for parts of bursts \citep[see also][]{Beloborodov+13,Li+19}. If comparing Figs. \ref{fig5:differentEpdis} and \ref{fig6:Ep-fluence} with Fig. \ref{fig7:Physical Ep}, one can conclude that those GRBs with higher $E_{\text p}>$ 100 keV are most probably interpreted by the Synchrotron radiations and the CPL model will be invalid for their $E_{\text p}$ estimations.

\section{Discussion and conclusions}
\label{sec:summary}

Regarding the $T_{90}$ distribution of \textit{Swift}/BAT bursts, how many components and where they are separated are two more and more controversial questions since the first research by \cite{Zhang+08}, where they analyzed the first four-year \textit{Swift} GRB data and pointed out that two types of GRBs divided at 2s are preferred. Meanwhile, the intrinsic $T_{90}$ was also found to be bimodally populated. The classification criteria are confirmed by our recent analysis of \textit{Swift}/BAT observations with a popular Bayesian Information Criterion (BIC) method if short GRBs with EE are excluded \citep{Zhang+16}. It is notable that we did not consider the spectral effects on the $T_{90}$ determination yet. Besides, the $T_{90}$ distribution is more or less biased by the background level, bin size, threshold and energy coverage of detector, sampling standard and so on. \cite{Bromberg+13} constructed an empirical model involving $T_{90}$ to distinguish the collapsar GRBs from the non-collapsar ones and they proposed to classify \textit{Swift} short and long bursts at 0.8$\pm$0.3 s physically. If so, our short GRBs with $T_{90}<$1 s are most probably originated from non-collapsars. Since the duration timescale is tightly connected with radiation region \citep{Zhang+07}, shorter $T_{90}$ implies these short GRBs might be produced from even smaller emitting radii, which should put further strict constraints on GRB theories.

\cite{Ghirlanda+04} obtained the mean $E_{\text p}$ values of BATSE GRBs to be 355$\pm$30 keV and 520$\pm$90 keV for short and long bursts, respectively, of which they are comparable in a sense. They insisted that it is the low energy power-law index in the BAND function causing the higher hardness ratio in short against long GRBs. Excitingly, we also find no differences between peak energies of $87^{+112}_{-49}$ keV for short bursts and $85^{+101}_{-46}$ keV for long ones even within narrower energy bands of \textit{Swift}/BAT. Consequently, one can conclude that the differences between short and long GRBs are less evident when the sample is restricted to a certain energy window. Moreover, our finding demonstrates that a significant fraction of short GRBs with lower $E_{\text p}$ does exist and could be an important extension to the low end of $E_{\text p}$. Lately, \cite{Begue+20} sorted Fermi/GBM short and long bursts and reported a correlation of $ T_{90}\propto E_{p}^{-0.35}$ that will become less correlative if \textit{Swift}/BAT short GRBs are included. Note that there is no obvious $T_{90}$-$E_{p}$ relation at all in our $E_{\text p}$-selected \textit{Swift}/BAT GRB sample.

Although the $E_{\text p}$ values of \textit{Swift}/BAT GRBs are on average lower than those of Fermi/GBM or CGRO/BATSE, this surprisingly does not affect the existence and consistency of $E_{\text p}$ in association with luminosity and isotropic energy in the higher $E_{\text p}$ region. In practice, \cite{Zhang+18} found that the power-law relations of $E_{p,i}\propto L_{\text p}^{0.44}$, $E_{p,i}\propto E_{iso}^{0.34}$ and $E_{p,i}\propto E_{\gamma}^{0.20}$ do exist in the \textit{Swift} GRB dominated samples and these empirical correlations are marginally consistent with some previous results. In addition, they noticed for the first time that the power-indexes are equal between short and long GRBs for any one of three energy correlations. These accordant results indicate that there is no considerable energetic evolution effect for these energy correlations. Moreover, some recent investigations show that pulse width and photon energy of not only CGRO/BATSE but also \textit{Swift}/BAT short GRBs are negatively related with a similar power-law form to long bursts, which is also found to be independent of the energy bands of detectors \citep{Li+2020a,Li+2020b}. 

To depict the observed GRB spectra, thermal and synchrotron/Comptonized radiation mechanisms are usually favoured in despite of other processes proposed especially for lower $E_{\text p}$ explanations. This is likely true when the low $E_{\text p}$ can not be described by the sole BB or a burst does not hold the thermal component at all. \cite{Lazzati+05} studied the spectra of 76 BATSE short bursts and found that about 75\% of the bursts have the spectra inconsistent with a BB-like form. However, previous radiation transfer simulations showed that that the Band function could be generated from a thermal origin and the resulting maximum $E_{\text p}$ will rise up to 3 MeV in the source frame \citep{Beloborodov+13}. It needs to point out that the high-redshift interpretation \citep{Heise+13} to lower $E_{\text p}$ may not be trustworthy because the $E_{\text p}$ is not correlated with redshift at least for our sample. Additionally, \cite{Katsukura+20} illustrated that the $E_{\text p}$ diversity can not be interpreted by the off-axis model and would be caused by some intrinsic mechanisms. Hopefully, the real formation mechanisms of the lower peak energies will be completely revealed theoretically and observationally in the new era of large telescopes.

 Based on the above investigations, we now summarize our results as follows. First, we have used a sample of 283 \textit{Swift}/BAT GRBs with well-measured $E_{\text p}$ to analyze their logarithmic $T_{90}$ distribution and found that the $T_{90}$ is still bimodally rather than triply distributed. The best fit with a two-gauss model gives two separated peaks at 0.21 s and 42.66 s divided by a boundary of $\sim$1.06 s that is consistent with 1.27 s \citep{Gruber+14} and $0.8\pm0.3$ s \citep{Bromberg+13} reported for Fermi/GBM and \textit{Swift}/BAT catalogs, respectively. Second, we find that the peak energies of \textit{Swift}/BAT GRBs are comparable with mean peak energies of $87^{+112}_{-49}$ and $85^{+101}_{-46}$ keV for short and long bursts, respectively. This is similar to what revealed for the BATSE short and long GRBs by \cite{Ghirlanda+04}. Thus the peak energy may not be a ideally representative parameter describing the spectral hardness. Third, the comparative studies of distinct methods for the $E_{\text p}$ estimates demonstrate that the Bayesian model and the BAND function can return more consistent $E_{\text p}$, while the frequently-used cut-off model will evidently overestimate the $E_{\text p}$ within higher energy scope, say $E_{\text p}>$100 keV. Fourth, we theoretically analyze the underlying $E_{\text p}$ formation mechanisms under condition that thermal, synchrotron radiation, and both of them (mixed radiations) are assumed to explain the $E_{\text {p,o}}$ diversity. Approximately, half sample of \textit{ Swift} GRBs including XRFs and most XRBs are located in the thermal-dominated regions, while another half sample of\textit{ Swift} GRBs with a majority of C-GRBs are mainly contributed by the non-thermal radiation components. Noticeably, there is only a small fraction of XRBs that can be interpreted by the mixture of the BB and the Synchrotron radiation emissions. Finally, we find a tight correlation between the $S_{\gamma}$ and the $E_{\text {p,o}}$ for the \textit{Swift}/BAT long GRBs to be $E_p\simeq [S_{\gamma}/(10^{-5} erg\ cm^{-2})]^{0.28}\times 117.5^{+44.7}_{-32.4}$ keV that might be applied as an indicator of peak energies if only the observed fluence is available.


\acknowledgments

We thank the referee for very constructive comments that have greatly improved the quality of our paper. This research has made use of the data supplied by the High Energy Astrophysics Science Archive Research Center (HEASARC) for the online \textit{Swift }catalog. It was partly supported by the Natural Science Foundations (20165660, ZR2018MA030, XKJJC201901 and 201909118).

\bibliographystyle{aasjournal}

\begin{thebibliography}{}
\expandafter\ifx\csname natexlab\endcsname\relax\def\natexlab#1{#1}\fi
\providecommand{\url}[1]{\href{#1}{#1}}
\providecommand{\dodoi}[1]{doi:~\href{http://doi.org/#1}{\nolinkurl{#1}}}
\providecommand{\doeprint}[1]{\href{http://ascl.net/#1}{\nolinkurl{http://ascl.net/#1}}}
\providecommand{\doarXiv}[1]{\href{https://arxiv.org/abs/#1}{\nolinkurl{https://arxiv.org/abs/#1}}}


\bibitem[{{Band} {et~al.}(1993)}]{Band+93}
{Band}, D., {Matteson}, J., \& {Ford}, L., et~al., 1993,\apj, 413, 281, \dodoi{10.1086/172995}

\bibitem[{{Barraud} {et al.}(2005)}]{Barraud+05}
{Barraud} C.,  {Daigne} F. \& {Mochkovitch} R., 2005, \aap, 440, 809, \dodoi{10.1051/0004-6361:20041572}

\bibitem[{{Beloborodov}(2013)}]{Beloborodov+13}
{Beloborodov}, A. M., 2013, \apj, 764, 157, \dodoi{10.1088/0004-637X/764/2/157 }


\bibitem[{{Berger}(2014)}]{Berger-14}
{Berger} E., 2014, \araa, 52, 43,  \dodoi{10.1146/annurev-astro-081913-035926}


\bibitem[{{Bosnjak} {et~al.}(2014)}]{Bosnjak+14}
{Bosnjak}, Z. \& Daigne, F., 2014, \apj, 568, 45, \dodoi{10.1051/0004-6361/201322341}


\bibitem[{{Bromberg} {et~al.}(2013)}]{Bromberg+13}
{Bromberg}, O., {Nakar}, E., \& {Piran}, T., Sari, R., 2013, \apj, 764, 179, \dodoi{10.1088/0004-637X/764/2/179}


\bibitem[{{Butler} {et~al.}(2007)}]{Butler+07}
{Butler}, N. R., {Kocevski}, D., \& {Bloom}, J. S., et al., 2007, \apj, 671, 656, \dodoi{ 10.1086/522492}


\bibitem[{{Chattopadhyay} {et~al.}(2007)}]{Chattopadhyay+07}
{Chattopadhyay}, T., { Misra}, R., \& {Chattopadhyay}, A. K., et~al., 2007, \apjl, 667, 1017, \dodoi{ 10.1086/520317}



\bibitem[{{Chattopadhyay} \& {Maitra}(2017)}]{Chattopadhyay+18}
{Chattopadhyay}, S., {Maitra}, R., 2017, \mnras, 469, 3374, \dodoi{ 10.1093/mnras/stx1024}

\bibitem[{{Daigne} \& {Mochkovitch}(2003)}]{Daigne+03}
{Daigne}, F., {Mochkovitch}, R., 2003, \mnras, 342, 587, \dodoi{10.1046/j.1365-8711.2003.06575.x}

\bibitem[{{Dereli-Begue} {et~al.}(2020)}]{Begue+20}
{Dereli-B\'{e}gu\'{e}}, H., {Pe'er}, A. \& {Ryde}, F., 2020, \apj, 897, 145,
\dodoi{10.3847/1538-4357/ab9a2d}

\bibitem[{{Dodelson}(2003)}]{Dodelson-03}Dodelson, S., 2003, Modern Cosmology
(Academic Press)

\bibitem[{{Ghirlanda} {et al.}(2004)}]{Ghirlanda+04}
{Ghirlanda}, G., {Ghisellini}, G. \& {Celotti}, A., 2004, \aap, 422, L55, \dodoi{10.1051/0004-6361:20048008}

\bibitem[{{Giannios}(2012)}]{Giannios+12}
{Giannios}, D., 2012, \mnras, 422, 3092, \dodoi{ 10.1111/j.1365-2966.2012.20825.x}


\bibitem[{{Gehrels} {et~al.}(2004)}]{Gehrels+04}
{Gehrels}, N., {Chincarini}, G., {Giommi}, P., et~al., 2004, \apj, 611, 1005, \dodoi{10.1086/422091}


\bibitem[{{Goldstein} {et~al.}(2010)}]{Goldstein+10}
{Goldstein}, A., {Preece}, R. D., {Briggs}, M. S., 2010,
\apj, 721, 132, \dodoi{10.1088/0004-637X/721/2/1329}



\bibitem[{{Goldstein} {et~al.}(2013)}]{Goldstein+13}
{Goldstein}, A., {Preece}, R. D., {Mallozzi}, R. S., et~al., 2013,
\apjs, 208, 21, \dodoi{ 10.1088/0067-0049/208/2/21}

\bibitem[{{Gruber} {et~al.}(2014)}]{Gruber+14}
{Gruber}, D., {Goldstein}, A., {Weller von Ahlefeld}, V., et~al., 2014,
\apjs, 211, 12, \dodoi{ 10.1088/0067-0049/211/1/12}


\bibitem[{{Heise}(2013)}]{Heise+13}
{Heise}, J., 2013, GAMMA-RAY BURST AND AFTERGLOW ASTRONOMY 2001: A Workshop Celebrating the First Year of the HETE Mission. AIP Conf. Proc., 662, 229, \dodoi{ 10.1063/1.1579346}


\bibitem[{{Horv\'{a}th} \& {T\'{o}th}(2016)}]{Horvath+16}
{Horv\'{a}th}, I., {T\'{o}th}, B. G., 2016,
\apss, 361, 155, \dodoi{ 10.1007/s10509-016-2748-6}


\bibitem[{{Klebesadel} {et~al.}(1973)}]{Klebesadel+73}
{Klebesadel}, R. W., {Strong}, I. B., \& {Olson}, R. A., 1973,
\apjl, 182, L85, \dodoi{ 10.1086/181225}


\bibitem[{{Katsukura} {et~al.}(2020)}]{Katsukura+20}
{Katsukura}, D., {Sakamoto}, T., \& {Tashiro}, M. S., 2020,
\apj, 889, 110, \dodoi{ 10.3847/1538-4357/ab6167}


\bibitem[{{Kouveliotou} {et~al.}(1993)}]{Kouveliotou+93}
{Kouveliotou} C. et~al,  1993, \apj, 413, 101, \dodoi{10.1086/186969}


\bibitem[{{Lazzati} {et al.}(2005)}]{Lazzati+05}
{Lazzati}, D., {Ghirlanda} G. \& {Ghisellini}, G., 2005, \mnras, 362, 8, \dodoi{10.1111/j.1745-3933.2005.00062.x}

\bibitem[{{Li} \& {Sari}(2008)}]{Li+08}
{Li}, C. \& {Sari} R., 2008, \apj, 677, 425, \dodoi{10.1086/527551}


\bibitem[{{Li}(2019)}]{Li+19}
{Li}, L., 2019, \apjs, 245, 7, \dodoi{10.3847/1538-4365/ab42de}

\bibitem[{Li} {et al.}(2020a)]{Li+2020a} {Li}, X. J., {Zhang}, Z. B., {Zhang}, C. T., et al., 2020a, \apj, 892, 113, \dodoi{10.3847/1538-4357/ab7a94}

\bibitem[{Li} {et al.}(2020b)]{Li+2020b} {Li}, X. J., {Zhang}, Z. B., {Zhang}, X. L., et al., 2020b, \apjs, submitted{}

\bibitem[{{Lien} {et al.}(2016)}]{Lien+16}
{Lien}, A., {Sakamoto}, T., {Barthelmy}, S. D., et al., 2016, \apj, 829, 7, \dodoi{10.3847/0004-637X/829/1/7}

\bibitem[{{M{\'e}sz{\'a}ros} (2006)}]{Meszaros-06}
{M{\'e}sz{\'a}ros} P., 2006, Rep. Prog. Phys., 69, 2259,\dodoi{10.1088/0034-4885/69/8/R01}


\bibitem[{{Nakar}(2007)}]{Nakar-07}
{Nakar} E.,  2007, Phys.Rept., 442, 166, \dodoi{10.1016/j.physrep.2007.02.005}
\bibitem[{{Norris} {et al.}(2000)}]{Norris+00}
{Norris} J. P., {Marani} G. F., {Bonnell} J. T., 2000, \apj, 534, 248, \dodoi{ 10.1086/308725}

\bibitem[{{Norris} \& {Bonnell}(2006)}]{Norris+06}
{Norris} J. P., {Bonnell} J. T., 2006, \apj, 643, 266, \dodoi{ 10.1086/502796}
\bibitem[{{Piran}(2004)}]{Piran-04}
{Piran} T.,  2004, Reviews of Modern Physics, 76, 1143, \dodoi{10.1103/RevModPhys.76.1143},

\bibitem[{{Rybicki}\& {Lightman}(1979)}]{Rybicki-79}Rybicki, G. B., \& Lightman, A. P., 1979, Radiative Processes in Astrophysics
(New York: Wiley-Interscience)

\bibitem[{{Ryde}(2004)}]{Ryde-04}
{Ryde} F., 2004, \apj, 614, 827, \dodoi{10.1086/423782}

\bibitem[{{Preece} {et~al.}(2000)}]{Preece+00}
{Preece} R. D.,  {Briggs} M. S.,   {Mallozzi} R. S., 2000, \apjs, 126, 19,  \dodoi{10.1086/313289}

\bibitem[{{Preece} {et al.}(2016)}]{Preece+16}
{Preece} R. D., {Goldstein} A., {Bhat} N., 2016, \apj, 812, 12,  \dodoi{10.3847/0004-637X/821/1/12}


\bibitem[{{Sakamoto} {et al.}(2004)}]{Sakamoto+04}
{Sakamoto} T.,  {Lamb} D. Q.,   {Graziani} C., 2004, \apj, 602, 875, \dodoi{10.1086/381232}

\bibitem[{{Sakamoto} {et~al.}(2005)}]{Sakamoto+05}
{Sakamoto} T., {Lamb} D. Q., {Kawai} N., et al., 2005, \apj, 629, 311, \dodoi{ 10.1086/431235}

\bibitem[{{Sakamoto} {et al.}(2008)}]{Sakamoto+08}
{Sakamoto} T., {Barthelmy} S. D., {Barbier} L., et al., 2008, \apjs, 175, 179, \dodoi{ 10.1086/523646}

\bibitem[{{Sakamoto} {et al.}(2009)}]{Sakamoto+09}
{Sakamoto} T., {Sato} G., {Barbier} L., et al., 2009, 693, 922, \apj, \dodoi{10.1088/0004-637X/693/1/922}

\bibitem[{{Sakamoto} {et al.}(2011)}]{Sakamoto+11}
{Sakamoto} T., {Barthelmy} S. D., {Baumgartner} W. H., et al., 2011, \apjs, 195, 2, \dodoi{10.1088/0067-0049/195/1/2}

\bibitem[{{Sari} {et~al.}(1998)}]{Sari+98}{Sari} R., {Piran} T., {Narayan} R., 1998, \apjl, 497, L17, \dodoi{10.1086/311269}

\bibitem[{{Svinkin} {et al.}(2016)}]{Svinkin+16}{Svinkin} D. S., {Frederiks} D. D., {Aptekar} R. L., 2016, \apjs, 224, 10, \dodoi{10.3847/0067-0049/224/1/10}


\bibitem[{{Tarnopolski}(2017)}]{Tarnopolski+17}{Tarnopolski} Mariusz, 2017, \mnras, 472, 4819, \dodoi{ 10.1093/mnras/stx2356 }

\bibitem[{{Tarnopolski}(2019a)}]{Tarnopolski+19a}{Tarnopolski} Mariusz, 2019a, \apj, 887, 97, \dodoi{ 10.3847/1538-4357/ab4fe6}

\bibitem[{{Tarnopolski}(2019b)}]{Tarnopolski+19b}{Tarnopolski} Mariusz, 2019b, \apj, 870, 105, \dodoi{ 10.3847/1538-4357/aaf1c5}

\bibitem[{{T\'{o}th} {et al.}(2019)}]{Toth+19}
{T\'{o}th}, B. G., {R\'{a}cz}, I. I., {Horv\'{a}th}, I., 2019, \mnras, 486, 4823, \dodoi{  10.1093/mnras/stz1188}


\bibitem[{{Yu} {et~al.}(2015)}]{Yu+15}
{Yu} H. F., {Greiner}, J., {van Eerten}, H., et~al., 2015, \aap, 573, 81, \dodoi{ 10.1051/0004-6361/201424858}

\bibitem[{{Zhang} \& {M{\'e}sz{\'a}ros}(2004)}]{Zhang-04}
{Zhang} B., {M{\'e}sz{\'a}ros}, P., 2004, International Journal of Modern Physics A, 19, 2385, \dodoi{ 10.1142/S0217751X0401746X}


\bibitem[{{Zhang} {et al.}(2009)}]{Zhang-09}{Zhang} B., {Zhang}, B. B., {Virgili}, F. J, et~al., 2009, \apj, 703, 1696, \dodoi{10.1088/0004-637X/703/2/1696}


\bibitem[{{Zhang} \& {Yan}(2011)}]{ZhangB-11}
{Zhang} B., {Yan}, H. R., 2011, \apj, 726, 90, \dodoi{10.1088/0004-637X/726/2/90}

 \bibitem[{{Zhang} {et al.}(2012)}]{ZhangB-12}
{Zhang} B., et al., 2012, \apjl, 758, L34, \dodoi{ 10.1088/2041-8205/758/2/L34}


\bibitem[{{Zhang} {et~al.}(2006)}]{Zhang+06}
{Zhang} Z.~B., et al., 2006, \mnras, 373, 729, \dodoi{ 10.1111/j.1365-2966.2006.11058.x}


\bibitem[{{Zhang} {et~al.}(2007)}]{Zhang+07}
{Zhang} Z.~B., {Xie} G. Z., {Deng} J. G., et al., 2007, AN, 328, 99, \dodoi{ 10.1002/asna.200610666}

\bibitem[{{Zhang} \& {Choi}(2008)}]{Zhang+08}
{Zhang} Z.~B., {Choi} C.~S., 2008, \aap, 484, 293, \dodoi{ 10.1051/0004-6361:20079210}


\bibitem[{{Zhang} {et al.}(2016)}]{Zhang+16}
{Zhang} Z.~B., {Yang} E. B., {Choi} C. S., et al., 2016, \mnras, 462, 3243, \dodoi{  10.1093/mnras/stw1835}

\bibitem[{{Zhang} {et~al.}(2018)}]{Zhang+18}
{Zhang} Z.~B., {Zhang} C. T., Zhao, Y. X., et al, 2018, \pasp, 130, 054202, \dodoi{10.1088/1538-3873/aaa6af}

 \bibitem[{{Zitouni} {et~al.}(2015)}]{Zitouni-15}
{Zitouni}, H.,  {Guessoum}, N., { Azzam}, W. J., {Mochkovitch}, R., 2015, \apss, 357, 7, \dodoi{10.1007/s10509-015-2311-x}

\bibitem[{{Zitouni} {et~ al.}(2018)}]{Zitouni-18}
{Zitouni}, H.,  {Guessoum}, N., {AlQassimi}, K. M., {Alaryani}, O., 2018, \apss, 363, 223, \dodoi{10.1007/s10509-018-3449-0}

\end{thebibliography}



\end{document}